\documentclass[intlimits]{amsart}
\usepackage[dvips]{graphicx}

\usepackage{bbm}
\usepackage{mathrsfs}
\usepackage{amssymb, amsbsy}

\newcommand{\abs}[1]{\left|#1\right|}           
\newcommand{\mat}[1]{\boldsymbol{#1}}           

\begin{document}

\title{Built-up structure criticality}
\author{Daniel Va\v sata${}^{1,2}$, Pavel Exner${}^{2,3}$ and Petr \v Seba${}^{2,4}$}
\maketitle
\begin{center}
\small{${}^1$ Department of Physics, Faculty of Nuclear Sciences and Physical Engineering\\
Czech Technical University in Prague, B\v rehov\'a 7, CZ-115 19, Prague, Czech Republic}\\
\small{${}^2$ Doppler Institute for Mathematical Physics and Applied Mathematics\\
Faculty of Nuclear Sciences and Physical Engineering\\
Czech Technical University in Prague, B\v rehov\'a 7, CZ-115 19, Prague, Czech Republic}\\
\small{${}^3$ Nuclear Physics Institute, Academy of Sciences of the Czech Republic\\
CZ-25068, \v Re\v z near Prague, Czech Republic}\\
\small{${}^4$ Faculty of Science, University of Hradec Kr\'alov\'e, V\'ita Nejedl\'eho 573\\
CZ-50002, Hradec Kr\'alov\'e, Czech Republic}\\

\end{center}

\begin{abstract}
The built-up land represents an important type of an overall
landscape. In this paper the built-up land structure in the largest
cities in the Czech Republic and some selected cities in the U.S.A.
is analyzed using the framework of statistical physics. We calculate
the variance of the total area and of the count of the   built-up
land plots contained inside discs of different radii. In both cases
the variance as a function of the disc radius follows a power law
with exponents that are comparable through different cities. The
study is based on the cadastral data from the Czech Republic and on
the building footprints from GIS data in the U.S.A.
\end{abstract}
\maketitle


\section{Introduction}
Urban land represents one of the most significant fingerprints of
human activity on the Earth. The creation and development of its
structure is influenced by cultural, sociological, economic,
political and other conditions. Despite the apparent complexity,
some simple universal properties and rules were found. The classic
example is the rank size distribution of cities firstly mentioned by
Auerbach (in \cite{Schweitzer-book}) and later discussed by Zipf
\cite{Zipf-Human-1949}. They claimed that if the cities are ranked
by the number of inhabitants, then the rank-size distribution
follows a power law with the exponent close to -1 (see also
\cite{rev2cl2}).

From the physical point of view it is interesting to study spatial
properties of the urban structure. Existing studies
\cite{Schweitzer-book, batty-fractal-cities, frank1, rev1cl1,
rev1cl2, rev2cl2, Pumain} focused especially on the fractal
structure and the related scaling-laws of urban clusters. The
analyzed pattern is usually given by some coarse-grained map of the
spatial population spread and the general claim is that the spatial
structure of large urban areas is influenced by certain long-range
spatial correlations. Motivated by those properties several urban
models were introduced \cite{Schweitzer-book, batty-fractal-cities,
rev1cl2, MakseNature, MaksePhysRewE, rev1cl4, rev2cl1}.

Our aim here is to study the urban structure in a different
representation and on much smaller scales. The analyzed data
consists of the exact positions of buildings in the cities
where we can expect the urbanization to be only slowly
varying with time. The vertical projection of buildings onto the
ground gives a built-up land pattern, that forms a natural
representation of the city since it is clearly visible and
identifiable. Built-up land is usually stable over short and medium
time periods as days, months and even years, except of rapid changes
during disasters. It is straightforward to expect that the built-up
land pattern is correlated with population spread since the human
live is strongly connected with buildings. The exact relation
however can be rather complicated as it depends on many additional
features like the whole 3D shape (capacity) or the usage (living,
working) of every building and can in general vary during the day
and over short time periods (weekends, vacations e.g.).

In this paper we show that the correlation and fluctuation
properties of such built-up land pattern in the  city centres are
similar as those of the critical systems in thermodynamics. This can
represent a connection between the urban and the critical systems.
It can be especially helpful when discussing the self-organized
criticality concept \cite{PhysRevA.38.364} that was for the urban
system introduced by Batty and Xie \cite{Batty98} with a fractal
dimension as the criticality indicator.

A connection to the critical systems (phase transitions) can
originate from the fact that economically the change of a land to a
built-up type represents a change of phase. The land acquires  an
additional property - a building.

\section{Critical phenomena}
Let us briefly recall the correlation and fluctuation properties
typical for thermodynamic systems near the critical point (e.g.
\cite{herbut, huang, landau}). For a further discussion on the
scaling properties in complex systems see \cite{rev3cl1}.

The most important property of the static spacial structure of a
critical system is its scaling invariance. In simple terms, if a
part of the system is magnified to the same size as the original
system, it is not possible to distinguish between the magnified part
and the original system.

In order to describe these features explicitly let us define a local order parameter $m$
as a quantity that solely describes the microscopic state of the system in one realization.
Thus $m(\mat{r})$ is the value of the parameter (e.g. density, local magnetization or boolean indicator of
some property occurrence) at the position $\mat{r} \in V$, where $V\subset \mathbbm{R}^2$ is the total volume occupied
by the 2-dimensional system living in the plane.

Suppose there exist a whole ensemble of realizations (different cities can be treated as different realizations) and
denote by $\langle ... \rangle$ the ensemble average.
We say that the system is homogeneous and isotropic in the volume $V$, if
\begin{equation}
\langle m(\mat{r})\rangle = \langle m(0)\rangle = m,\qquad \forall\mat{r} \in V.
\end{equation}
This means that the mean value of the local order parameter is
independent of position in the volume. From now we assume the system
to be homogeneous and isotropic.

The spatial properties of the order parameter distribution can be
described by the two-point correlation function defined as
\begin{equation}
G(\mat{r}_1, \mat{r}_2) = \bigg\langle \Big(m(\mat{r}_1) - \langle m(\mat{r}_1)\rangle\Big)
\Big(m(\mat{r}_2) - \langle m(\mat{r}_2)\rangle\Big)\bigg\rangle.
\label{corfun}
\end{equation}
This is under homogeneity and isotropy assumptions simplified to
\begin{equation}
G(\mat{r}_1, \mat{r}_2) = G(\mat{r}_2 - \mat{r}_1) = G(r) = \langle m(\mat{r})m(0)\rangle -m^2,
\end{equation}
where $\mat{r} = \mat{r}_2 - \mat{r}_1$ and $r = \abs{\mat{r}}$.
Therefore the correlation function depends only on the relative
distance $\abs{\mat{r}} = \abs{\mat{r}_2 - \mat{r}_1}$ of the two
points $\mat{r}_1$ and $\mat{r}_2$.

The scaling assumption for systems at the critical point can be written in the following form\footnote{
Using notation: $f(x) \sim g(x)$ $\Leftrightarrow$ $\lim_{x \rightarrow \infty}\frac{f(x)}{g(x)} = c$
and  $f(x) \propto g(x)$ $\Leftrightarrow$ $\frac{f(x)}{g(x)} = c,\ \forall x$.}:
\begin{equation}
G(r) \sim \frac{1}{r^{\eta}},\quad 0 < \eta < 2,
\label{corsca}
\end{equation}
Index $\eta$ appearing in the exponent of the power law part of
$G(r)$ is called the anomalous dimension. For systems outside the
critical point the correlation function decays with increasing $r$
much faster (usually exponentially).

The definition of the order parameter can be extended to systems
composed of point particles. Here, the empirical density function
$\rho$ is taken as the local order parameter. For particles located
at points $\mat{r}_1,\ \mat{r}_2,\ \mat{r}_3,\ \mat{r}_4,\ ... \in
\mathbbm{R}^2$ it is given by
\begin{equation}
\rho(\mat{r}) = \sum_{i=1}^\infty \delta(\mat{r} - \mat{r}_i).
\label{densitydef}
\end{equation}
It is well known \cite{landau} that the correlation function for such density can be decomposed to
\begin{equation}
G(\mat{r}_2 - \mat{r}_1) = \rho_0\delta(\mat{r}_2 - \mat{r}_1) + \mathcal{G}(\mat{r}_2 - \mat{r}_1),
\label{cordiag}
\end{equation}
where $\mathcal{G}(\mat{r}_2 - \mat{r}_1)$ is the non-diagonal part
of the form \eqref{corsca} defined for $r = \abs{\mat{r}_2 -
\mat{r}_1}
>0$. The difference to the ordinary order parameter is thus only in the
diagonal $\delta$ therm which of course doesn't influence the
character of the divergence in the vicinity of the critical point.
\subsection{Parameter variance in discs}
An useful tool to analyze the experimental data is the variance of
the parameter value inside discs. For the parameter $m(\mat{r})$
with homogeneous and isotropic distribution $\langle
m(\mat{r})\rangle = m$ the cumulative value of the parameter in the
disc of a radius $R$ is given by
\begin{equation}
M(R) = \int_{S(R)} m(\mat{r}) d\mat{r},
\label{MR0def}
\end{equation}
where the disc is the set $S(R) \equiv \{\mat{r}\in
\mathbbm{R}^2|\abs{\mat{r}}<R\}$ with a volume $\abs{S(R)}$. The
centre of the disc is not important due to the homogeneity of the
parameter distribution. The parameter variance is defined
\cite{PhysRevD.65.083523} as
\begin{equation}
\sigma^2(R) = \langle M(R)^2\rangle - \langle M(R)\rangle^2,
\label{paramvar}
\end{equation}
where
\begin{equation}
\langle M(R) \rangle = \int_{S(R)} \langle m(\mat{r})\rangle d\mat{r} = m \abs{S(R)},
\label{MRdef}
\end{equation}
and
\begin{equation}
\langle M(R)^2 \rangle = \int_{S(R)}\int_{S(R)} \langle m(\mat{r}_1) m(\mat{r}_2)\rangle d\mat{r}_1 d\mat{r}_2.
\label{MR2def}
\end{equation}

Using the definition \eqref{corfun} of the two-point correlation
function and the fluctuation-dissipation theorem \cite{huang}, one
obtains two different asymptotic relations for $\sigma^2(R)$:

Outside of the critical point the following relation holds:
\begin{equation}
\sigma^2(R) \sim  \langle M(R)\rangle \propto R^2,\qquad R \gg 1.
\label{subpois}
\end{equation}

A different situation arises when the system is approaching the
critical point. Spatial correlations in this region are long-ranged
and the correlation function is dominated by the power-law decay
\eqref{corsca}. This gives
\begin{equation}
\sigma^2(R) \sim \langle M(R)\rangle^{2 - \frac{\eta}{2}} \propto R^{4 - \eta},\qquad  1 \ll R \ll \xi,
\label{critbehave}
\end{equation}
where $\xi$ is the correlation length of the system that goes to
infinity as the system approaches the critical point. The
correlation length can be understood as the range of interactions.

Thus, in order to determine the criticality of a thermodynamic
system, one can study its fluctuations. If the dependence of
$\sigma^2(R)$ on $\langle M(R)\rangle$ follows the power law with
exponent larger than 1, than the system is close to the critical
point. The breakdown of the power-law \eqref{critbehave} for very
large $R$ is connected with reaching the correlation length. However
because of the complicated relation between the correlation function
$G(r)$ and $\sigma^2(R)$, a more appropriate way to determine the
correlation length $\xi$ is a direct study of the correlation
function and breakdown of the scaling form \eqref{corsca}.

\section{Data analysis}
In this part we show how the method from previous section can be
applied to the built-up land pattern. The analyzed data consist of
two different datasets:

\medskip
\noindent\emph{Cadastral records in the Czech Republic}

\begin{figure}[t]
\begin{minipage}[t]{.5\linewidth}
\includegraphics[width=0.9\textwidth]{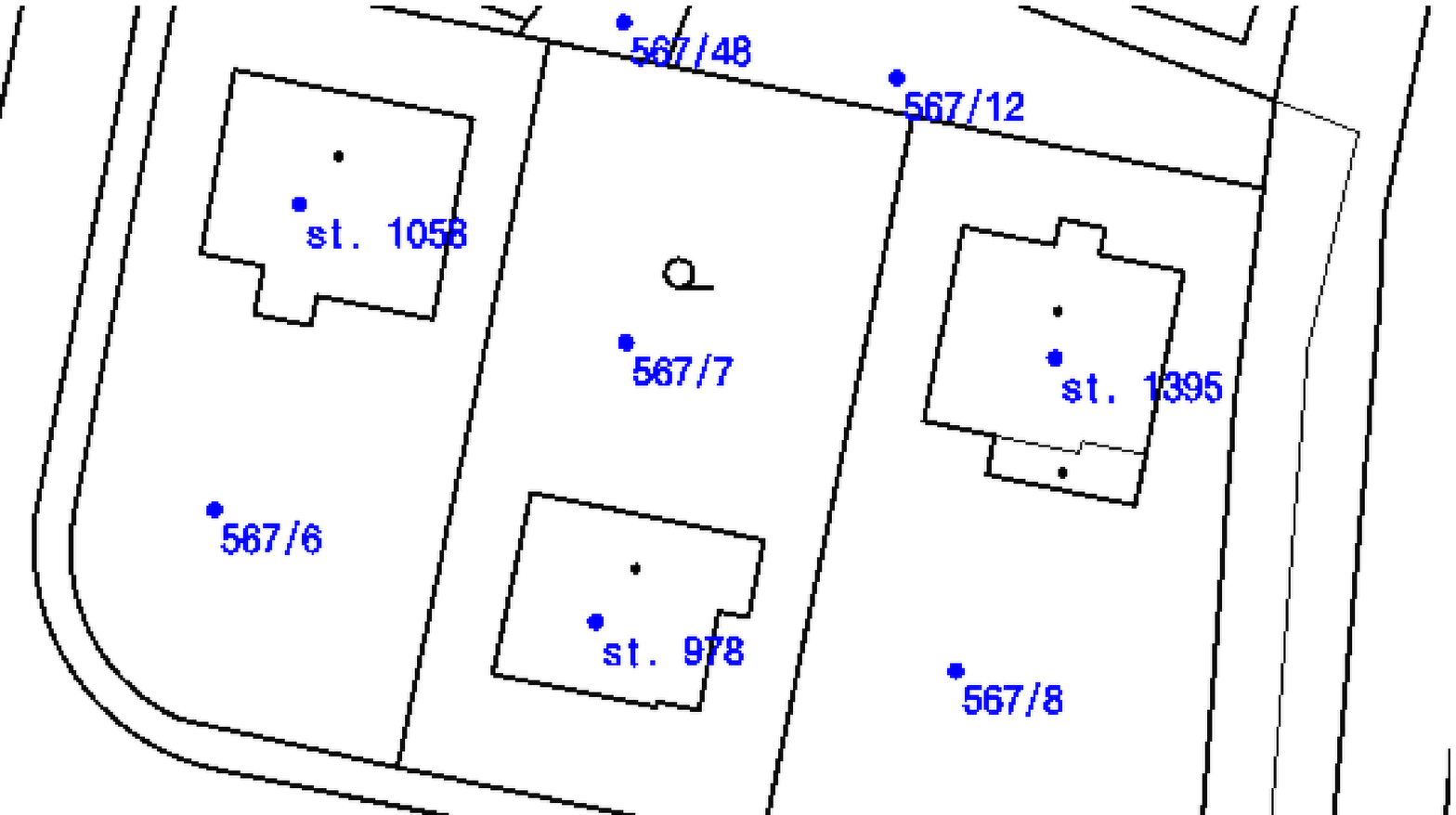}
{\noindent
\caption{\small Example of the Czech Republic dataset. Blue points
represent the definition points of the parcels. The polygon data
clearly showing the real shape of the parcels are not accessible to
our analysis.} \label{crdata}
}
\end{minipage}\hfill
\begin{minipage}[t]{.48\linewidth}
\includegraphics[width=0.9\textwidth]{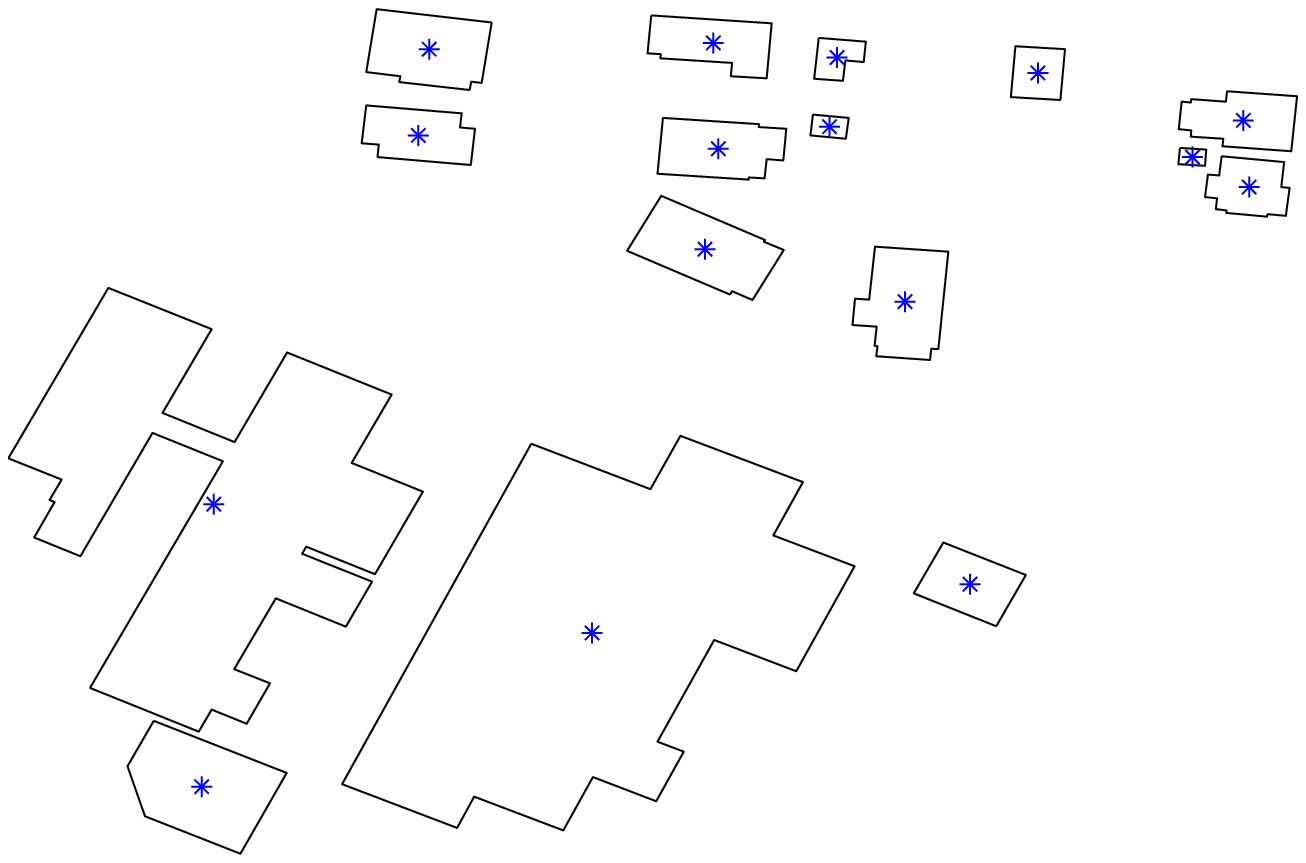}
\caption{\small Example of the U.S.A. dataset. Building footprints are defined as polygons. Centroids of polygons are
plotted as blue points.}
\label{usadata}
\end{minipage}
\end{figure}

The first dataset is formed the by cadastral records stored by COSMC
(Czech Office for Surveying Mapping and Cadastre). In general
cadastral records contain information about the fractalization of
the overall landscape into the smallest unique pieces of land -
\emph{the land plots (parcels)}. In the Czech Republic, every land
plot $i$ is characterised by its definition point $\mat{r}_i$, exact
geodetic shape, size (acreage) $\lambda_i$, type of land and the
ownership information. Our data contains all information except of
the exact shape for all land plots in the CR. The only geodetic
information is thus the definition point of the land parcel that
is the point located approximately at the centroid of the parcel.
The example of this data is shown in figure \ref{crdata}.

Since our interest is in the built-up structure, we restrict our
attention to the built-up land plots only (built-up land plot =
a building on it).

\medskip
\noindent\emph{Building footprints in the U.S.A.}

The second part of our data, the building footprints, are part of
the GIS data in the ESRI Shapefile format available for few U.S.A.
cities on the Internet (see section Resources for links). The
building footprints are represented by polygons in the plane. Those
polygons reflect the vertical projection of the overlaying buildings
to the ground. Visualisation of a small part of the data is shown on
the figure \ref{usadata}.
\subsection{Representation}
In order to analyse those different datasets we use two different
but straightforward representations:

\medskip
\noindent\emph{Point representation} is given by the definition
points (CR dataset) resp. centroids of polygons (U.S.A. dataset).
For every city this gives a set of points $\{\mat{r}_i, i\in I\}$.
The order parameter that characterizes such point pattern is given
by the singular point density $\rho(\mat{r})$ given by
\eqref{densitydef}. The parameter variance $\sigma^2(R)$ means the
variance of the number of points inside  discs.

\begin{figure}[ht]
\begin{center}
\includegraphics[width=0.7\textwidth]{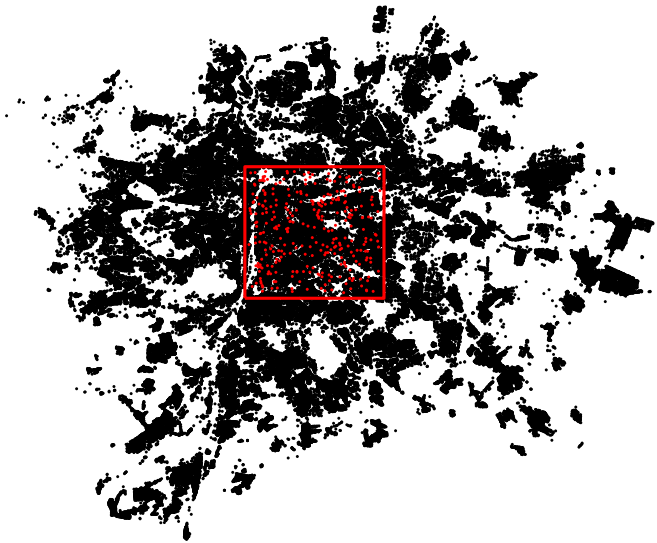}
\caption{The point representation of Prague built-up land plots.
The analyzed area is bounded by the red rectangle. The randomly
positioned centres $\mat{o}_j$ of 200 discs are plotted as red
points.} \label{figpointcalc}
\end{center}
\end{figure}

The estimation of $\sigma^2(R)$ for a given radius $R$ is done in
the following way: Inside the investigated part of the city $A
\subset \mathbbm{R}^2$ we uniformly and randomly choose centres
$\mat{o}_j$ of $N$, $N >> 1$ (usually $N = 2000$) discs, so that
every disc is a subset of $A$, $S(\mat{o}_j,R) \equiv \mat{o}_j +
S(R) \subset A$. For every disc $S(\mat{o}_j,R)$ the number of inner
points $N_j(R)$ is calculated,
\begin{equation}
N_j(R) = \int_{S(\mat{o}_j,R)} \rho(\mat{r}) d\mat{r}.
\end{equation}
The mean value is then estimated by
\begin{equation}
\langle N(R)\rangle = \frac{1}{N}\sum_{j} N_j(R)
\end{equation}
and the variance by
\begin{equation}
\sigma^2(R) = \frac{1}{N - 1}\sum_{j} \Big(N_j(R) - \langle N(R) \rangle\Big)^2.
\label{paramvarest}
\end{equation}
The example of a working area selection for Prague is shown in
figure \ref{figpointcalc}.

\medskip
\noindent\emph{Set representation} better reflects the existing
structure of the built-up land. Let assume that we work only with
the second dataset. Therefore the build-up land is given by
non-intersecting polygons $\{p_i \subset \mathbbm{R}^2, i\in I\}$.
Such set  can be represented as a subset of a plane given by boolean
order parameter
\begin{equation}
m(\mat{r}) = \left\{
\begin{array}{rll}
1 & \mathrm{building\ at\ } \mat{r},& (\exists i\in I, \mat{r} \in p_i)\\
0 & \mathrm{othervise}, & (\forall i\in I,  \mat{r} \notin p_i)\\
\end{array}
\right.
\end{equation}
The parameter variance $\sigma^2(R)$ can be estimated by use of \eqref{MR0def}.

If we want to deal with unknown shapes of the built-up parcels in
the Czech case, the reasonable way is to approximate them by discs
with the same  area. Such approximation however cannot lead to the
same (set) definition of the order parameter, because we generally
cannot avoid the discs to overlap. This actually is not a problem
and we can introduce the equivalent process of estimating
$\sigma^2(R)$ that can be easily extended to the overlapping
approximation.

Let us now suppose that our built-up parcels are  given as a set
$\{p_i \subset \mathbbm{R}^2, i\in I\}$ but the non-intersecting
property is generally not valid. This set is easily  obtained for
both datasets. In the Czech case $p_i$ is the circle of the same
area $\lambda_i$ as the $i$-th land plot located at its definition
point $\mat{r}_i$,
\begin{equation}
p_i = \left\{\mat{r}\in \mathbbm{R}^2|\abs{\mat{r} - \mat{r}_i}< \sqrt{\lambda_i/\pi}\right\}.
\end{equation}
In the U.S.A. $p_i$ remains the polygon of the $i$-th building.

The estimation of $\sigma^2(R)$ for a given radius $R$ is then done
as follows: Inside the studied part of the city $A \subset
\mathbbm{R}^2$ we uniformly and randomly choose centres $\mat{o}_j$
of $N$, $N >> 1$ discs, so that every disc is contained in $A$,
$S(\mat{o}_j,R) \equiv \mat{o}_j + S(R) \subset A$. For every disc
$S(\mat{o}_j,R)$ the built-up area $M_j(R)$ inside it is calculated
by the relation
\begin{equation}
M_j(R) = \sum_{i\in I} \lambda\left(p_i \cap S(\mat{o}_j,R)\right),
\end{equation}
where $\lambda(A)$ stands for the area of the set $A$. In other
words, we accumulate the area of every intersection of the
land plot $p_i$ with the given disc $S(\mat{o}_j,R)$ over all
land plots.

\begin{figure}[ht]
\begin{center}
\includegraphics[width=0.7\textwidth]{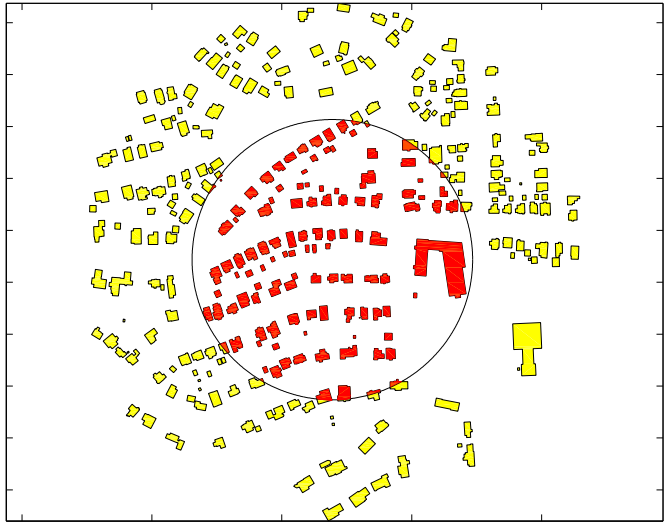}
\caption{The example of a built-up area calculation for a set
representation of the U.S.A. dataset. The area $M_j(R)$ of polygons
inside the disc $S(\mat{o}_j,R)$ is marked as red.}
\label{figsetcalc}
\end{center}
\end{figure}

The process of calculating the area inside the disk $S(\mat{o}_j,R)$
for polygons is depicted in figure \ref{figsetcalc}.

The calculation of $M_j(R)$  for the polygons (U.S.A. dataset) is
exact. On the other hand the disc approximation  gives only an
approximate result. The error is however not large and it is easy to
show that the effective error  is inversely proportional $R$
\begin{equation}
\frac{\widehat{M_j(R)} - M_j(R)}{\widehat{M_j(R)}} \sim R^{-1},
\end{equation}
with increasing $R$. Here $\widehat{M_j(R)}$ is the correct built-up
area acreage inside the $j$-th disc. This is because for $R$ much
larger than the typical parcel radius, deviations are produced only
in the vicinity of the large disc boundary.

The mean value is  estimated by
\begin{equation}
\langle M(R)\rangle = \frac{1}{N}\sum_{j} M_j(R)
\end{equation}
and the variance by
\begin{equation}
\sigma^2(R) = \frac{1}{N - 1}\sum_{j} \Big(M_j(R) - \langle M(R) \rangle\Big)^2.
\label{paramvarest2}
\end{equation}

The estimations for the both representations are based on the
assumption of self-averaging property \cite{Sornette,
PhysRevLett.77.3700}. It means that a sufficiently large sample is a
good representative of the whole ensemble. In our case however, the
size of the sample is limited by the size of the city centre. By the
city centre we mean the area around the city core (central
'plateau'), where the built-up density is virtually constant
\cite{Pumain}. This part of the city does not participate in a
process of massive urbanization in contrary to the edges of the
city. It thus represent a structure in a steady state (slowly
varying). This does not exclude some local urbanization changes that
are always presented. In order to speak about a virtually uniform
density it is also necessary to omit certain locations like lakes or
hilly places where it is impossible to construct a building from the
analysis. Otherwise the fluctuations will increase.

\section{Results}
We analyzed the 6 largest cities in the Czech Republic and 6 cities
in the U.S.A. In the centre of every city we calculated both the
point number variance in discs (point representation) and the
built-up area variance in discs (set representation) for different
values of the disc radius $R$. The diameter of the city centre for a
typical large Czech city is about 4 km. This size puts limitation on
the maximal radius $R$ of discs in order to obtain reasonable
statistics. Together with the fact that the power law dependence, if
present, can be theoretically reached for $R\gg 1$, we decided to
study the fluctuations inside the region $200\ \mathrm{m}\ \lesssim
R \lesssim 1000\ \mathrm{m}$.

\begin{figure}[ht]
\begin{center}
\includegraphics[width=0.9\textwidth]{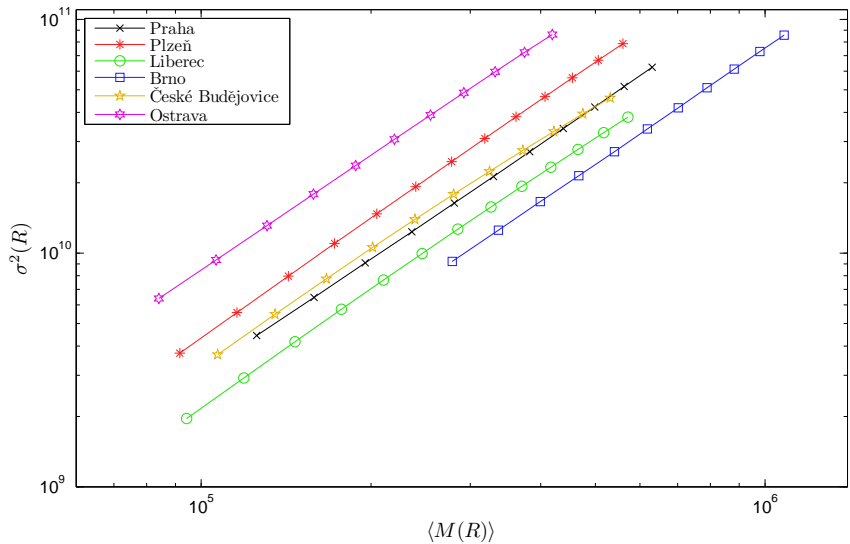}
\caption{Dependencies of $\sigma^2(R)$ for the set representation on $\langle M(R)\rangle$ for different
cities in the Czech republic.}
\label{fig1-1}
\end{center}
\end{figure}
\begin{figure}[t]
\begin{center}
\includegraphics[width=0.9\textwidth]{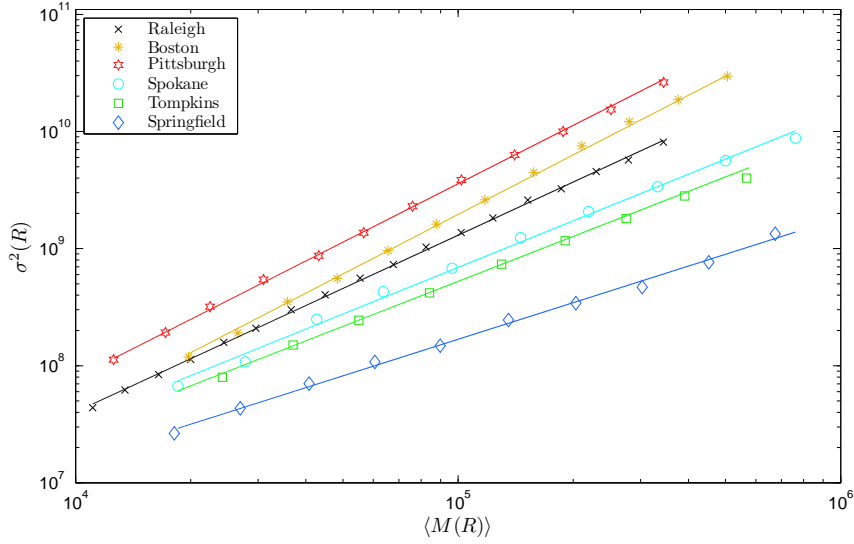}
\caption{Dependencies of $\sigma^2(R)$ for the set representation (polygon) on $\langle M(R)\rangle$ for different
cities in the U.S.A.}
\label{fig1-2}
\end{center}
\end{figure}
The obtained dependencies of $\sigma^2(R)$ on $\langle M(R)\rangle$
for the set representation are shown on figures \ref{fig1-1} and
\ref{fig1-2}. It is clearly visible that  for the set representation
the result follows power law (inside the analysed radius range). The
same behaviour is  valid also for the point representation. Thus in
the studied range the fluctuations behave as
\begin{equation}
\sigma^2(R) \propto \langle M(R)\rangle^{\alpha},\quad \alpha = 2 - \frac{\eta}{2}.
\label{alpharel}
\end{equation}
We determine the values of the exponent $\alpha$ by performing the linear regression on the logarithm relation
\begin{equation}
\log(\sigma^2(R)) = \alpha \log(\langle M(R)\rangle) + \beta.
\label{linreg}
\end{equation}
The dependency of $\langle M(R)\rangle$ on $R$ resp. $\sigma^2(R)$ on $R$
follows the power law with exponent 2 resp. $2\alpha$ as predicted by the homogeneity assumption resp.
the relation \eqref{critbehave}.

\begin{table}[t]
\caption{Exponents $\alpha$ according to the power law dependence
\eqref{alpharel} of $\sigma^2(R)$ on $\langle M(R)\rangle$. In the
2-nd and 5-th column are values of the exponent for the point
representation. The 3-rd and 6-th columns contain values for the set
representation with circular approximation of the built-up units
(for U.S.A. dataset - circles with the origin in the centroid and
the same size as actual polygon). The 7-th column contains values
for the set representation using the actual known polygonal shape of
the building footprints.} {\begin{tabular}{|c|c|c|c|c|c|c|} \hline
City & Points & Area&City & Points & \multicolumn{2}{|c|}{Area}\\
&& circle &&& \multicolumn{1}{|c}{circle} & \multicolumn{1}{c|}{polygon}\\
\hline
Praha & 1.47 & 1.64 & Raleigh & 1.73 & 1.58 & 1.58\\
Plze\v n & 1.61 & 1.69& Pittsburgh & 1.62 & 1.62 & 1.62\\
Liberec & 1.54 & 1.65 & Boston & 1.69 & 1.68 & 1.69\\
Brno & 1.40 & 1.65& Spokane & 1.69 & 1.57 & 1.59\\
\v Cesk\' e Bud\v ejovice & 1.50 & 1.58&Tompkins & 1.75 & 1.57 & 1.59\\
Ostrava & 1.54 & 1.62&Springfield &1.52 & 1.30 &1.30\\
\hline
\end{tabular}}
\label{tab1}
\end{table}
The summary of resulting exponents $\alpha$ for studied cities is
given in the table \ref{tab1}. As follows from \eqref{subpois},
$\alpha = 1$ stands for the system that is outside of the critical
region, e.g. randomly positioned particles. One can see that this is
not the case for the built-up land pattern.

The values of the exponent for the built-up land pattern are for
both representations much larger than 1. In the case of a point
pattern the average value of the exponent is $\overline{\alpha}_p =
1.60$. We can see a systematic difference between the Czech cities
(lower values) and the American ones (larger values).

\begin{figure}[ht]
\begin{center}
\includegraphics[width=0.6\textwidth]{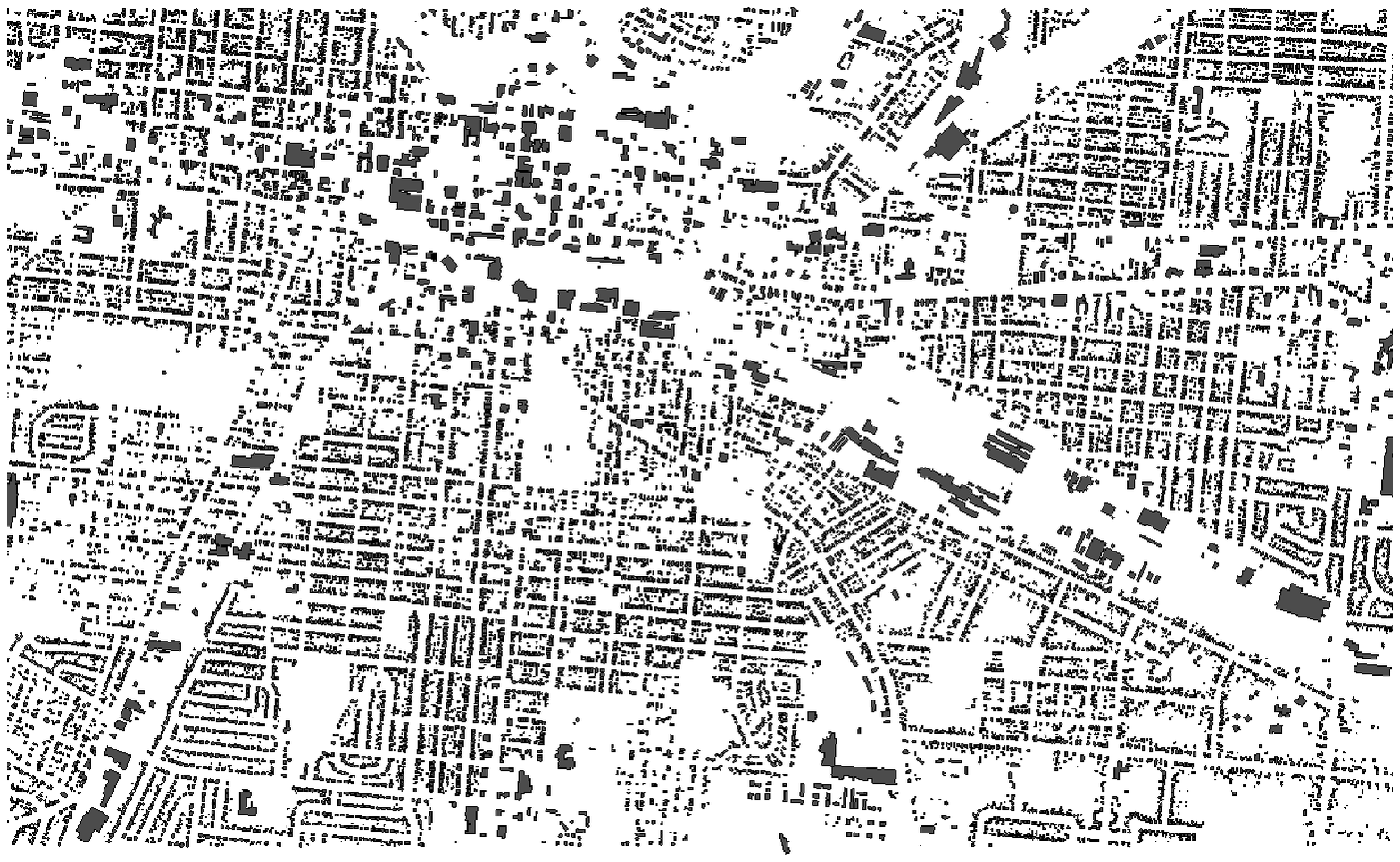}
\caption{The constrained structure of Springfield. The built-up land is plotted in its exact polygonal representation.} 
\label{fig4}
\end{center}
\end{figure}

More interesting results arise for the set representation. There is
not a clear systematic difference in this case between the Czech
republic and the U.S.A. The average value of the exponent is
$\overline{\alpha}_a = 1.62$ with much lower fluctuations around
this value. The only significant deviation in the power law
coefficient is represented by the City of Springfield (Clark
County). Such a result can be explained by its constrained
lattice-like structure (see figure \ref{fig4}). Because the
fluctuations are influenced by the local inhomogeneities of the
pattern, the lattice-like structure produce a more homogeneous
distribution (with lower fluctuations) of the built-up land than for
the other more "organic" cities \cite{batty-fractal-cities}.
Theoretically in the case of an exactly rigid structure with
identical buildings placed at the vertices of a perfect lattice the
fluctuations will increase and the coefficient $\alpha < 1$. For
further information on this so-called superhomogeneous distribution
see \cite{PhysRevD.65.083523}.

Comparing the values of the exponent $\alpha$ in the 6-th and 7-th
column of the table \ref{tab1} we can conclude that the
approximation of an unknown parcel shape by a circle does not
produce significant error.

\begin{figure}[ht]
\begin{center}
\includegraphics[width=0.8\textwidth]{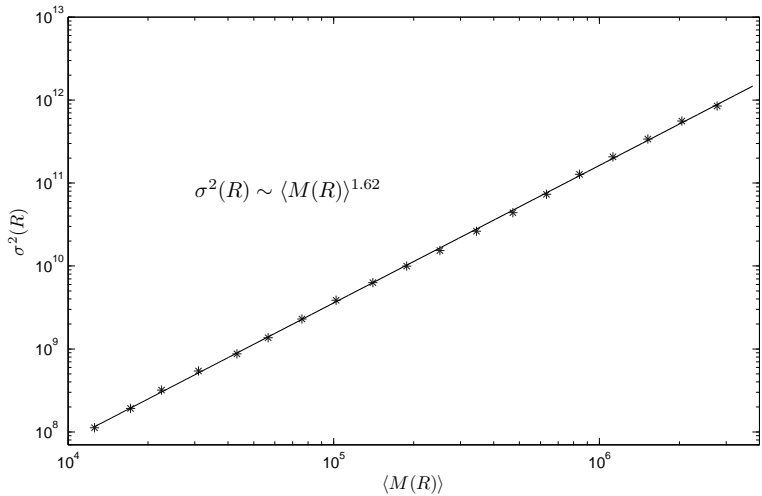}
\caption{Fluctuation dependence in Pittsburgh for a set representation
(polygons) in the region $R \in (200, 3000)$ m. Straight line
represents linear regression fit according to the logarithm relation
\eqref{linreg}.} \label{figPitt}
\end{center}
\end{figure}

Let us now make a short comment on the analysed range of $R$. In our
analysis we used the interval $(200\ \mathrm{m}, 1000\ \mathrm{m})$.
For some large cities in the U.S.A. we were eventually able to
increase the upper bound to 3 km, because the analysed city centre
area can be taken larger than for Czech cities that are much
smaller. The power-law dependence remains unchanged in the extended
range (see figure \ref{figPitt} for Pittsburgh in the set
representation).

Some further extension of the region can be also calculated but the
assumption about the homogeneity (constant density) of the
distribution is then obviously not valid since the density decreases
with the increasing distance from the city centre. This  influences
the presented estimations of $\sigma^2(R)$ because the
self-averaging cannot be applied. So even if the correlations may be
evaluated on much larger distances, it is not proper to use this
method outside of the theoretical homogeneous region.

\begin{figure}[ht]
\begin{center}
\includegraphics[width=0.8\textwidth]{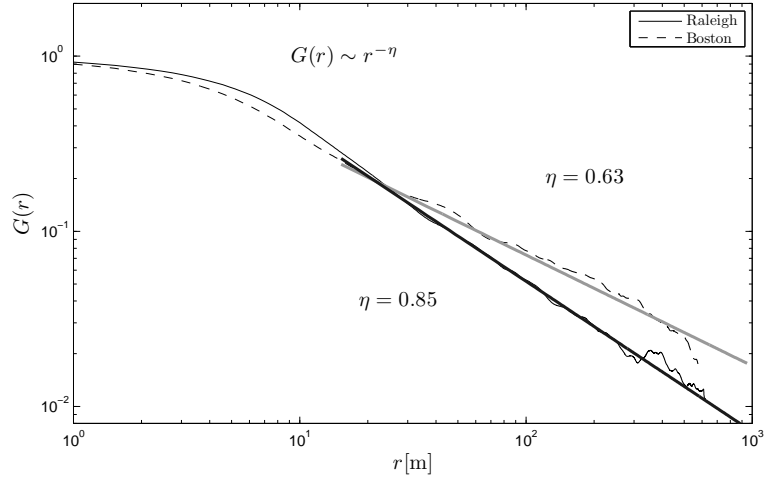}
\caption{The correlation function for Raleigh and Boston. Straight lines represent the
linear regression fit of correlation function tails. 
Expected power-law tail coefficients $\eta$ calculated from the respective values of $\alpha$ in table \ref{tab1} 
(according to \eqref{alpharel}) 
are $\tilde{\eta} = 0.84$ for Raleigh,
resp. $\tilde{\eta} = 0.62$ for Boston.} \label{fig2}
\end{center}
\end{figure}

For the polygon representation of the built-up land in the U.S.A. it
is also possible to estimate the correlation function directly.
Results for Raleigh and Boston are shown in figure \ref{fig2}.
The correlation function decay clearly follows a power law and the
exponents are consistent with the values of $\alpha$ from table
\ref{tab1} and the relation \eqref{alpharel}. The same consistency
holds for the rest of the studied cities.

\begin{figure}[ht]
\begin{center}
\includegraphics[width=0.6\textwidth]{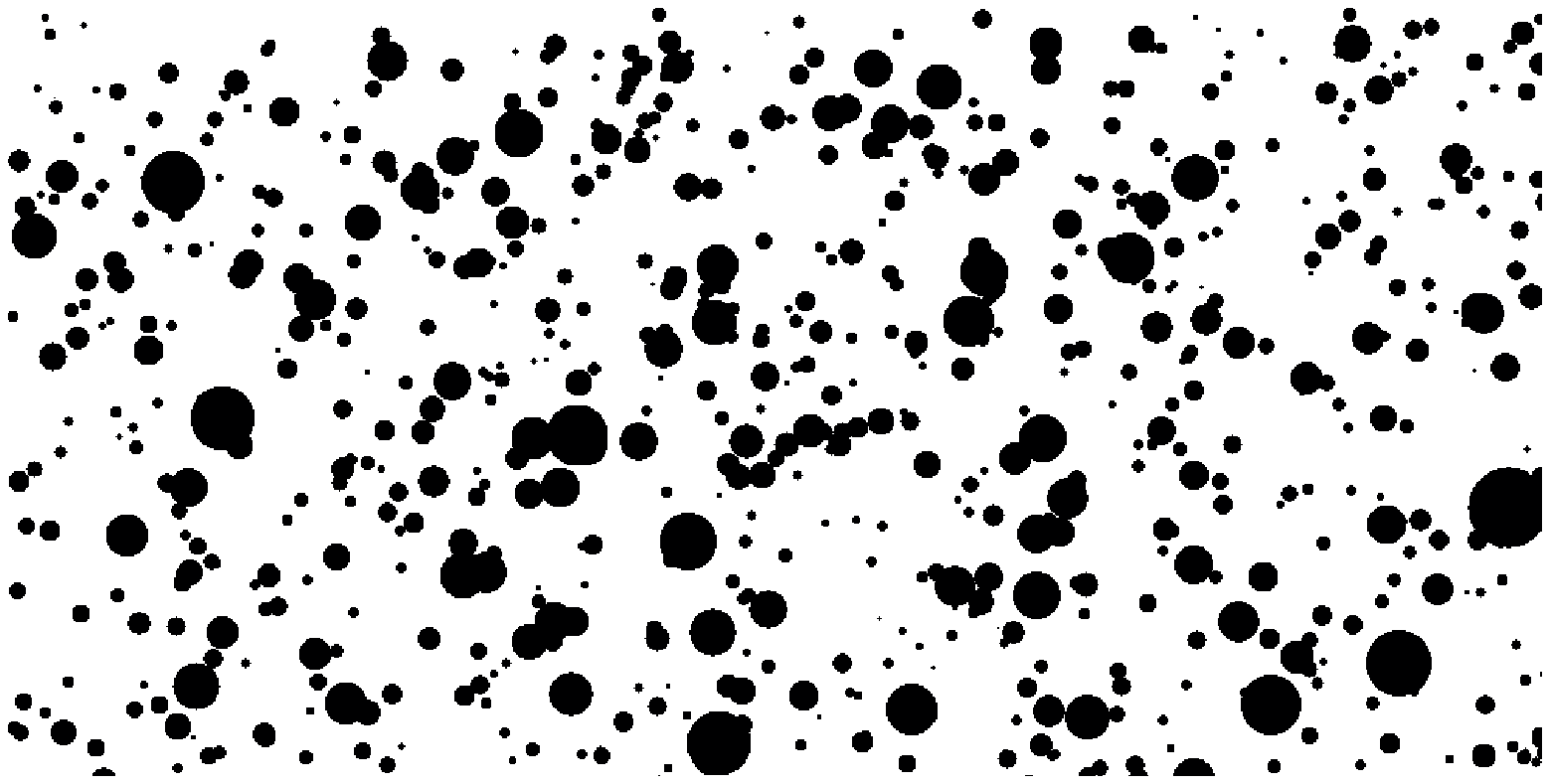}
\caption{Part of a randomly generated city.} \label{figrandcity}
\end{center}
\end{figure}

\begin{figure}[ht]
\begin{center}
\includegraphics[width=0.8\textwidth]{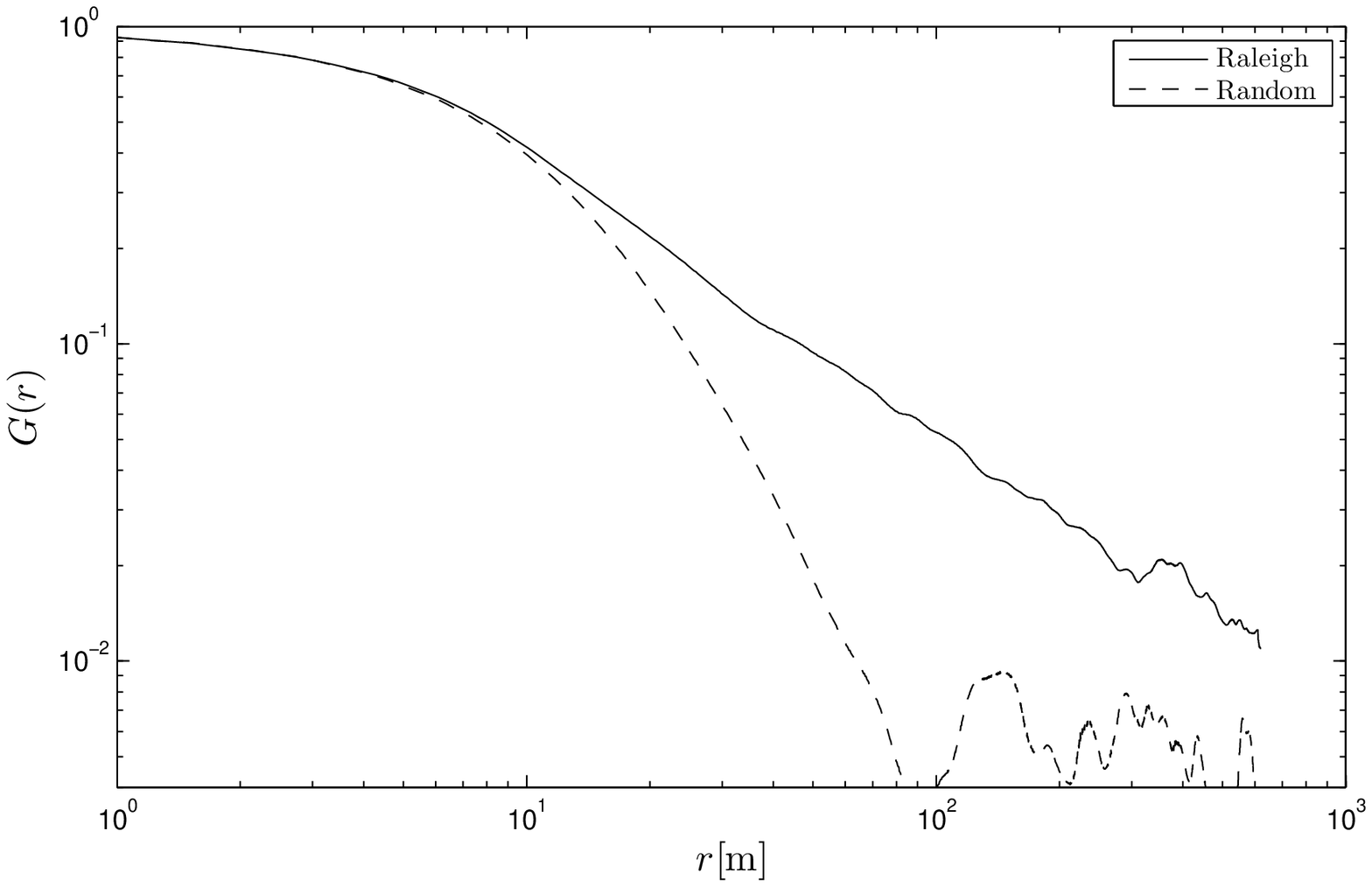}
\caption{The correlation function for randomly generated pattern
compared to the correlation function for Raleigh.} \label{fig3}
\end{center}
\end{figure}

Finally we generated an artificial city by positioning the centres
of the buildings randomly. Buildings were approximated by circles
with a size distributed with the same distribution as for a real
city. The part of this artificial city is shown in figure
\ref{figrandcity}.
The correlation function for such pattern compared with the
correlation function for Raleigh is shown in figure \ref{fig3}.
The fluctuation properties of such pattern are consistent with the
predictions given by \eqref{subpois}.

\section{Conclusion}
The study shows that dependence of fluctuations of the size of the
build up area on its mean value  follows a power law. Moreover the
set representation of the plots seems to lead to  more universal
results. The values of the exponent $\alpha$ in the relation
$\sigma^2(R) \sim \langle M(R)\rangle^{\alpha},$ for different
cities (except of Springfield) are all very close to the value
$\alpha = 1.62$. The different results for Springfield can be
explained by the strongly constrained lattice-like structure of that
city.

We can conclude that the inner urban area structure is correlated
with a long-ranged power-law dependence. The power-law exponent
seems to be independent of the particular city. Such an observation
is interesting and the possible connection between the urban area
correlations and the correlations in critical systems may be useful
for the development and verification of further urban models. The
fact that the inner (quasi-stable) part of the city has certain city
independent properties supports the hypothesis of the self-organized
criticality in the urban systems \cite{Batty98}. 
However we are not able to study the process of the
self-organization in details because our data do not reflect the
dynamics of city growth, as we have only one time snapshot for
each city.

\section*{Acknowledgments}
The research was supported by the Czech Ministry of Education, Youth
and Sports within the project LC06002, by the Grant Agency of the Czech Republic within the project No. 202/08/H072 and
by the project No. SGS10/211/OHK4/2T/14 of the Czech Technical University in Prague.
We are indebted to Helena \v{S}andov\'{a}
and Petr Sou\v{c}ek from the \emph{Czech Office for
Surveying, Mapping and Cadastre} for the help with acquiring the data.
We also thank to anonymous reviewers for valuable remarks.
\section*{Resources}
The building footprints for U.S.A. cities were obtained from the following web sites:
\begin{itemize}
\item Boston - \texttt{http://www.mass.gov/mgis/database.htm}
\item Raleigh - \texttt{http://www.wakegov.com/gis/default.htm}
\item Pittsburgh - \texttt{http://www.alleghenycounty.us/dcs/gis.aspx}
\item Spokane - \texttt{http://www.spokanecity.org/services/gis/}
\item Tompkins - \texttt{http://cugir.mannlib.cornell.edu/index.jsp}
\item Springfield (Clark County) - \texttt{http://gis.clark.wa.gov/gishome/}
\end{itemize}

\bibliographystyle{unsrt}
\bibliography{bib-crit}

\end{document}